# EXPERIMENTAL RESULTS AND ANALYSIS FROM THE 11T Nb$_3$Sn DS DIPOLE*

G. Chlachidze, I. Novitski, A.V. Zlobin, Fermilab, IL 60510, USA
B. Auchmann, M. Karppinen, CERN, Geneva, Switzerland

*Abstract*

FNAL and CERN are developing a 5.5-m-long twin-aperture Nb$_3$Sn dipole suitable for installation in the LHC. A 2-m-long single-aperture demonstrator dipole with 60 mm bore, a nominal field of 11 T at the LHC nominal current of 11.85 kA and 20% margin has been developed and tested. This paper presents the results of quench protection analysis and protection heater study for the Nb$_3$Sn demonstrator dipole. Extrapolations of the results for long magnet and operation in LHC are also presented.

## INTRODUCTION

The expected upgrade of the LHC collimation system foresees installation of additional collimators in the dispersion suppressor (DS) regions around points 2, 3, 7 and high-luminosity IRs in points 1 and 5 [1]. The space needed for the collimators could be provided by replacing 15-m-long 8.33 T Nb-Ti LHC main dipoles with shorter 11 T Nb$_3$Sn dipoles compatible with the LHC lattice and main systems [2]. CERN and FNAL have started a joint R&D program with the goal of building a 5.5-m-long twin-aperture Nb$_3$Sn dipole suitable for installation in the LHC [3]. The program started with the design [4], construction and test [5] of a 2-m-long 60 mm bore single-aperture demonstrator magnet.

Due to large stored energy (a factor of 1.5 larger than in the Nb-Ti LHC main dipoles) the protection of the 11 T Nb$_3$Sn dipoles in case of a quench is a challenging problem. As in all accelerator magnets including LHC main dipoles, it will be provided with dedicated protection heaters installed in the coil to spread the stored electromagnetic energy over larger coil volume and thus reduce its maximum temperature and electrical voltage to ground.

Heater position plays an important role in magnet protection. The traditional position of protection heaters in accelerator magnets is the outer surface of the coil outer layer (OL), used practically in all present accelerator magnets including the LHC main dipoles [6]. It provides excellent mechanical contact between the heaters and the coil, and allows adequate coil electrical insulation from ground. However, coil volume directly heated by the protection heaters is limited to ~50% of the total coil volume in this design.

To increase the coil volume affected by the protection heaters, they could be placed both on the inner and outer surfaces of the two-layer coil or inside the coil between the inner and outer layers. Installation of the protection heaters in the high field areas should also increase their efficiency. The inner-layer heaters were used in D20 [7] and in LARP LQS and HQ models [8, 9]. The inter-layer protection heaters were used in the first Nb-Ti MQXB short models (HGQ) [10] and in the first FNAL Nb$_3$Sn model (HFDA01) [11]. However, both these approaches have some difficulties. The inner-layer heaters add an additional thermal barrier between the coil and liquid helium in the annular channel, reducing the coil cooling conditions. Moreover, the mechanical contact between the heaters and the coil in this case is weak and could easily be destroyed during the magnet assembly, cooling down, or operation. Partial heater separation was observed in LARP quadrupoles after testing in superfluid helium at 1.9 K [8]. The inter-layer heaters have good mechanical contact with both coils but they require significant electrical reinforcement of the coil inter-layer insulation to withstand the high voltages which may lead to significant reduction of their efficiency. They could also be easily damaged during the Nb$_3$Sn coil reaction, magnet assembly, and operation. Due to the above-mentioned difficulties both these approaches have not been used yet in magnets operating at accelerators. That is why the quench protection development for 11 T Nb$_3$Sn dipoles has started with the traditional outer-layer protection heaters.

This paper describes the design and parameters of the protection heaters used in the 2-m-long demonstrator dipole, and presents the first experimental data and results of analysis of quench protection studies. Results are extrapolated to a 5.5-m-long magnet and operation in the LHC.

## MAGNET AND PROTECTION HEATER DESIGNS

Details of the 11 T demonstrator dipole design are reported in [4, 5]. The two-layer coils consist of 56 turns - 22 in the inner layer and 34 in the outer layer. Each coil is wound using 40 strand Rutherford cable [12] insulated with two layers of 0.075 mm thick E-glass tape. The cable is made of 0.7 mm diameter Nb$_3$Sn RRP-108/127 strand with a nominal $J_c$(12 T, 4.2 K) of 2750 A/mm$^2$ (without self-field correction), a copper fraction of 0.53, and RRR above 60 [13].

The coils are surrounded by multilayer ground insulation made of Kapton, stainless steel protection shells, and laminated stainless steel collars. The collared coil is installed inside a two-piece iron yoke clamped with two aluminum clamps and stainless steel shells. In the longitudinal direction the magnet is constrained with two thick stainless steel end plates.

* Work supported by Fermi Research Alliance, LLC, under contract No. DE-AC02-07CH11359 with the U.S. Department of Energy and European Commission under FP7 project HiLumi LHC, GA no. 284404.

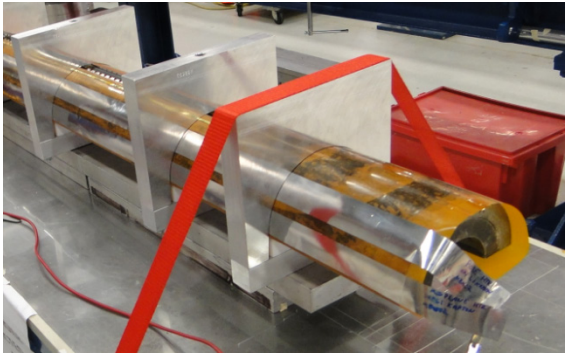

Figure 1: Two heater strips on one side of the coil.

Quench heaters are placed between the ground insulation layers of Kapton. The first Kapton layer, bonded to the coil outer surface, is 0.114 mm thick including the thin adhesive layer. All the remaining layers without an adhesive layer are 0.127 mm thick. The magnet quench protection heaters are composed of 0.025 mm thick and 2.108 m long stainless steel strips, 21 mm wide at the mid-plane low-field (LF) blocks and 26 mm wide at the high-field (HF) pole blocks. Two heater strips on one side of the coil are shown in Fig.1. The resistance at 300 K of HF and LF strips is 0.87 $\Omega$/m and 1.06 $\Omega$/m, respectively.

Two strips connected in series are inserted between the ground insulation layers on the outer surface of the coil blocks. The ground insulation design and protection heater position are shown in Fig. 2. Thickness of the insulation between the protection heaters and the coil is an important parameter for the heater efficiency and its electrical insulation from coil and ground. To find the optimal value for heater insulation satisfying the contradictory requirements two protection heaters were tested in the same coil. Each coil has two protection heaters marked as PH-1L and PH-2L. PH-1L is installed between the 1st and 2nd Kapton layers on one side of the coil and PH-2L - between the 2nd and 3rd Kapton layers on the opposite side.

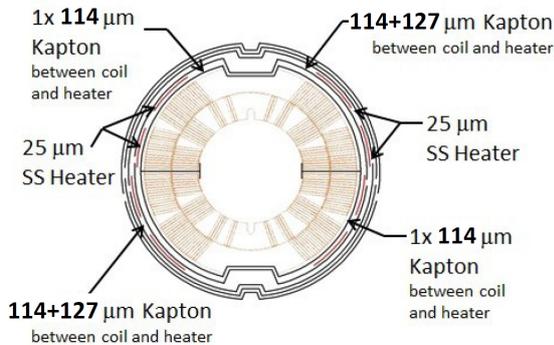

Figure 2: Ground insulation and protection heater position.

The corresponding protection heaters on each coil are connected in parallel forming two parallel heater circuits. The connection scheme of protection heaters in the 11 T dipole demonstrator is shown in Fig. 3. Each pair of protection heaters covers 31 turns (15 in the mid-plane and 16 in the pole block) per quadrant or ~56% of the total outer coil surface, or 28% of the total coil volume. The resistance of each protection heater measured at room temperature is ~5.9 $\Omega$ and ~4.2 $\Omega$ at 4.5 K.

Due to difference in width of heater strips (Fig. 1) the peak power density dissipated in the LF (mid-plane block) and HF (pole block) areas are also different. The peak power density in the low field area is more than in the high field area by about 50%.

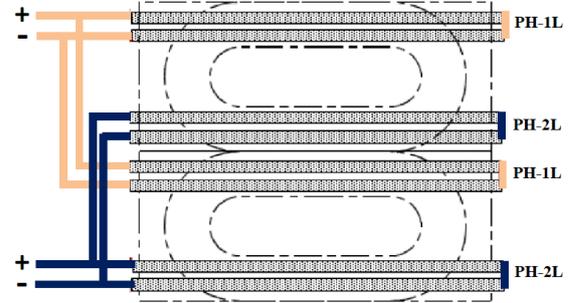

Figure 3: Heater connection scheme.

## QUENCH PROTECTION PARAMETERS

The quench protection parameters of the single-aperture 11 T $Nb_3Sn$ dipole at the LHC nominal current of 11.85 kA are summarized in Table 1. Table 2 shows the strand and cable parameters used in quench protection analysis.

## QUENCH PROTECTION ANALYSIS

### Coil Maximum Temperature and Quench Integral Limit

The maximum coil temperature $T_{max}$ after a quench in adiabatic conditions is determined by the equation:

$$\int_0^\infty I^2(t)dt = \lambda \cdot S^2 \cdot \int_{T_q}^{T_{max}} \frac{C(T)}{\rho(B,T)} dT \qquad (1)$$

where $I(t)$ is the current decay after a quench (A); $T_q$ is the conductor quench temperature (K); $S$ is the cross-section of the insulated cable (m$^2$); $\lambda$ is fraction of Cu in the insulated cable cross-section; $C(T)$ is the average volumetric specific heat of the insulated cable (J K$^{-1}$ m$^{-3}$); $\rho(B,T)$ is the cable resistivity ($\Omega$ m).

Table 1: Demonstrator dipole quench protection parameters

| Parameter | Value |
|---|---|
| Effective magnet length (m) | 1.7 |
| Number of turns per coil ($N_{turn}$/coil) | 56 |
| Nominal current (kA) | 11.85 |
| Current density in Cu stabilizer (kA/mm$^2$) | 1.362 |
| Inductance at $I_{nom}$ (mH/m) | 6.04 |
| Stored energy at $I_{nom}$ (kJ/m) | 424 |
| Energy density W/V$_{coil}$ (MJ/m$^3$) | 85.9 |
| Maximum quench field (T) | 13.4 |
| Critical quench current (kA) | 15.0 |
| Maximum stored energy (kJ/m) | 680 |

Table 2: Strand and cable parameters

| Parameter | Value |
|---|---|
| Cable width (mm) | 14.85 |
| Cable mid thickness (mm) | 1.307 |
| Strand diameter (mm) | 0.7 |
| Number of strands | 40 |
| Cu/SC ratio | 1.11 |
| Insulation thickness (mm) | 0.1 |
| Total cable area (mm$^2$) | 22.7 |
| Total strand area (mm$^2$) | 15.4 |
| Cu area (mm$^2$) | 8.08 |
| Non-Cu area (mm$^2$) | 7.31 |
| Insulation area (mm$^2$) | 3.27 |
| Void area filled with epoxy (mm$^2$) | 4.01 |
| Cu RRR | 100 |

The dependence of $T_{max}$ on the value of quench integral ($QI$) calculated for the demonstrator dipole cable insulated with E-glass tape and impregnated with epoxy for two values of the external magnetic field corresponding to the maximum and minimum fields in the coil is shown in Fig. 4. The thermal properties of the cable insulation (epoxy impregnated E-glass) were represented by G-10. Calculations were performed independently at FNAL and CERN using different databases for material properties. A good agreement of the results was obtained. Large effect of the magnetic field on the coil temperature is seen in Fig. 4. However, due to the current and field decay during a quench its effect on turn heating in the coil is smaller as shown in Fig. 5 where the magnetic field decay from $B_{max}$ to 0 is taken into account.

To keep the cable temperature during a quench below 400 K, the quench integral has to be less than 19-21 MIITs (10$^6$ A$^2$·s). This criterion for a maximum cable temperature (still under discussion) is currently considered as an acceptable limit for Nb$_3$Sn accelerator magnets [14].

*Protection delay budget*

The maximum value of the quench integral in the turn where the quench originated is determined by the equation:

$$\int_0^\infty I^2(t)dt = I_0^2 \tau_D + \int_{\tau_D}^\infty I^2(t)\,dt, \quad (2)$$

where $I_o$ is the magnet current when the quench started; $\tau_D$ is the total delay time including the quench detection, protection switch operation, and heater delay time; and $I(t)$ is the current decay in the magnet after the protection heaters were fired.

Protection heater parameters such as heater delay time (the time between the heater ignition and the start of quench development in the coil) and coil volume under the protection heaters as well as quench propagation velocity in the coil provide significant impact on $\tau_D$ and $I(t)$ in equation (2) and thus on the value of the maximum temperature in the quench origin area.

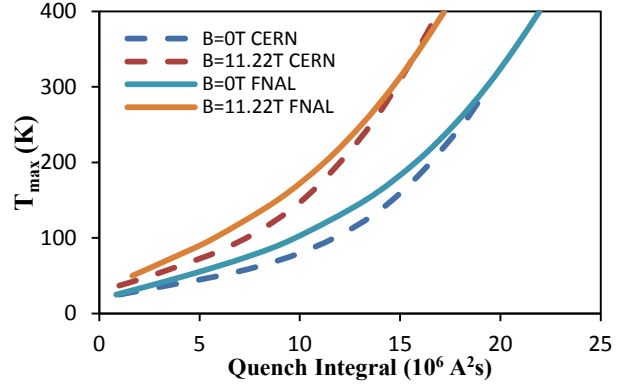

Figure 4: Cable maximum temperature $T_{max}$ vs. Quench Integral $QI$ for the insulated and epoxy-impregnated cable (strand RRR=100).

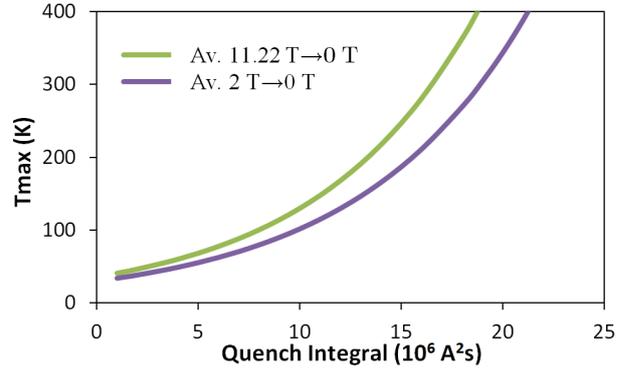

Figure 5: Cable maximum temperature $T_{max}$ vs. quench integral $QI$ for the insulated and epoxy-impregnated cable (strand RRR=100) corrected on the magnetic field decay in the IL pole turns ($B_{max}$=11.22 T) and the OL mid-plane turns ($B_{max}$=2 T).

The time budget $\tau_{budget}$ for $\tau_D$ (including the heater delay) is defined by the formula

$$\tau_{budget} = \frac{QI_{max} - QI_{decay}}{I_0^2}, \quad (3)$$

where the maximum quench integral $QI_{max}$ is calculated using (1) for the maximum allowed coil temperature of 400 K; $QI_{decay}$ is the quench integral accumulated during the current decay; and $I_0$ is the magnet quench current.

The $QI_{decay}$ could be estimated using formula (1) if the coil average maximum temperature under quench heaters $T^{PH}_{max}$ is known. This temperature was calculated assuming that all the turns under the protection heaters quench simultaneously and the magnet stored energy is dissipated only in these turns

$$\frac{W(I_o)}{l} \cong N_{qt} fS \int_{T_q}^{T^{PH}_{max}} C(T)\,dT, \quad (4)$$

where $W(I_0)/l$ is the stored energy per magnet unit length (J/m); $N_{qt}$ is the number of turns quenched by quench heaters; $f$ is the number of quench heaters used in each coil (1 or 2).

The average maximum coil temperature under the heaters vs. magnet current is shown in Fig. 6. The longitudinal and transverse quench propagation is not considered in these calculations. As it follows from the plot, at the nominal operation current 11.85 kA the coil maximum temperature under the heaters is less than 250 K, even with one operation heater circuit. $T^{PH}_{max}$ is an important parameter which defines also the coil stress due to coil expansion inside the cold structure.

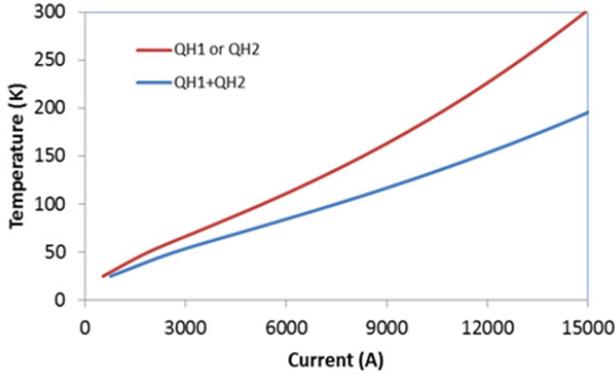

Figure 6: The average maximum coil temperature under the heater vs. magnet current for one and two protection heater circuits.

The calculated delay budget $\tau_{budget}$ for the inner-layer turns of the 11 T Nb$_3$Sn dipole vs. magnet current normalized to its short sample limit (SSL) is shown in Fig. 7 for protection with one and two heater circuits. The delay budget reduces with the magnet current reaching its minimum at the nominal operation current. For operation with two protection heaters the delay budget at $I_{nom}$ (80% of SSL) is 50 ms and for one heater only 25 ms. Delay budgets in the case of quench development in the coil outer layer are larger due to the lower magnetic field: 30-50 ms for one PH and more than 200 ms for two PHs respectively.

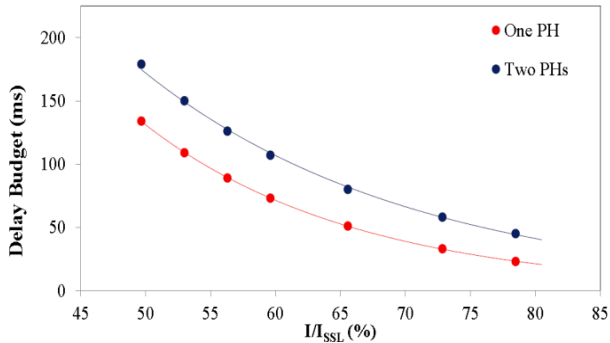

Figure 7: Calculated delay budget for the 11 T dipole vs. normalized magnet current.

### Quench and heat propagation

The analysis described above does not consider the longitudinal and transverse quench propagation in coil nor the heat transfer inside the coil and between the coil and the magnet support structure. These effects increase the effective coil volume involved in the energy dissipation as well as dissipate some fraction of the stored energy outside the coil reducing the maximum temperature in the quench origin area and under the quench heaters. Consequently, the delay budget will also increase.

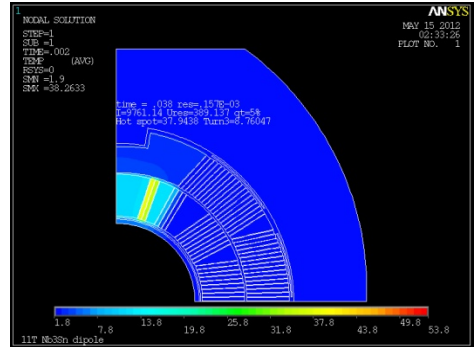

Figure 8: Temperature profile in the demonstrator magnet after 38 ms from the inner-layer pole turn quench.

The effect of the transverse heat propagation was analyzed using a 2D quench simulation code based on ANSYS [15]. Figure 8 shows the temperature profile in the demonstrator magnet after 38 ms from a quench at the nominal current of 11.85 kA in the inner-layer pole turn. It can be seen that the coil pole blocks and wedges are involved in the quench process absorbing a part of the dissipated heat and thus reducing the maximum temperature of quenched turn. Based on simulations the turn-to-turn propagation time is very short, less than 10 ms [16].

Figure 9 shows the temperature profile in the cross-section of the demonstrator dipole after 48, 96 and 552 ms from the heater induced quench at the coil initial current of 11.85 kA.

After ~50 ms from the protection heater discharge the quench starts in the outer-layer HF pole block. Then, in less than 100 ms, the quench propagates to the inner layer through the interlayer insulation. The outer-layer coil reaches its temperature of 150-213 K (compare with the average value of 150 K for QH1+QH2 in Fig. 6) after 550 ms from the heater ignition. As in the previous case, efficient heat transfer from the heater to the coil outer layer, from the outer-layer to inner-layer turns and other coil components helps to spread and absorb the magnet stored energy [16].

The results of the described quench analysis were further studied and experimentally verified during the quench protection studies in the 11 T demonstrator dipole [17].

## EXPERIMENTAL STUDIES

The 11 T demonstrator dipole was tested at FNAL Vertical Magnet Test Facility [18] in June 2012.

### Coil instrumentation

The coils were instrumented with voltage taps for the quench detection and localization. The voltage tap scheme for one of the coils is shown in Fig. 10. Voltage taps in pole turn allow measuring quench propagation velocity in the case of spontaneous quenches in this area. Voltage taps on each current block provide the quench propagation time between these blocks. In the next coils, spot heaters and more voltage taps will be added in coil mid-plane and pole areas to measure the quench propagation speed and turn heating after quench.

A series of tests was performed to evaluate the efficiency of the heaters with different insulation (PH-1L and PH-2L) and the ability to quench the coil with a reasonably short delay time. Heater delay time was defined as the time between the heater ignition and the start of quench development in the coil. For each test, a pair of heaters with a specific insulation was fired while another pair of heaters were used for the magnet protection along with the stored energy extraction system. Due to limited quench performance of the magnet [5], heater tests were performed only at currents up to 65% of the estimated short sample limit (SSL). The energy extraction circuit delay was 1 ms for all heater tests except for the radial quench propagation study, during which the extraction dump was delayed for 120 ms.

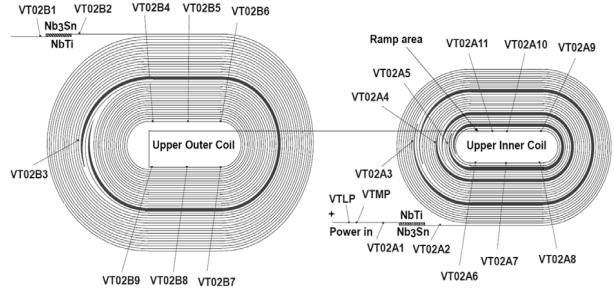

Figure 10: Voltage tap scheme in the 11 T demonstrator dipole coil.

*Protection heater delay*

Heater delay at a different SSL ratio ($I/I_{SSL}$) measured both at 4.5 K and 1.9 K is shown in Fig. 11 for the average heater power of 25 W/cm$^2$. Measured heater delay time is compared in Fig. 11 with the estimated delay budget presented in Fig. 7. Extrapolation of the measurement data to the nominal operation current (80% of the SSL) gives ~25 ms and ~40 ms heater delay time for PH-1L and PH-2L respectively. The corresponding extrapolated values at the injection current (5% of SSL) are ~420 ms and ~2000 ms.

The data in Fig. 11 show that the heater delay time is practically same at 4.5 K and 1.9 K temperatures, but it strongly depends on the heater insulation thickness. The dependence of the heater delay time on Kapton insulation thickness between the heater and the coil for the 11 T demonstrator dipole and some other Nb$_3$Sn coils used in LARP TQ and HQ models [8] are summarized in Fig. 12.

The measured heater delay time for PH-2L heaters with double Kapton layers of insulation itself is longer than the total available delay budget at all curents. The PH-1L heaters in the regular case, when both heaters are used for coil protection, provide ~25 ms margin with respect to the total delay budget which allows for necessary delays in the quench detection and circuit operation. However, in the case of only one heater operation (redundant case) this margin disappears. More time margin could be achieved by reducing the insulation thickness between the coil and heater, or increasing the peak dissipated power density.

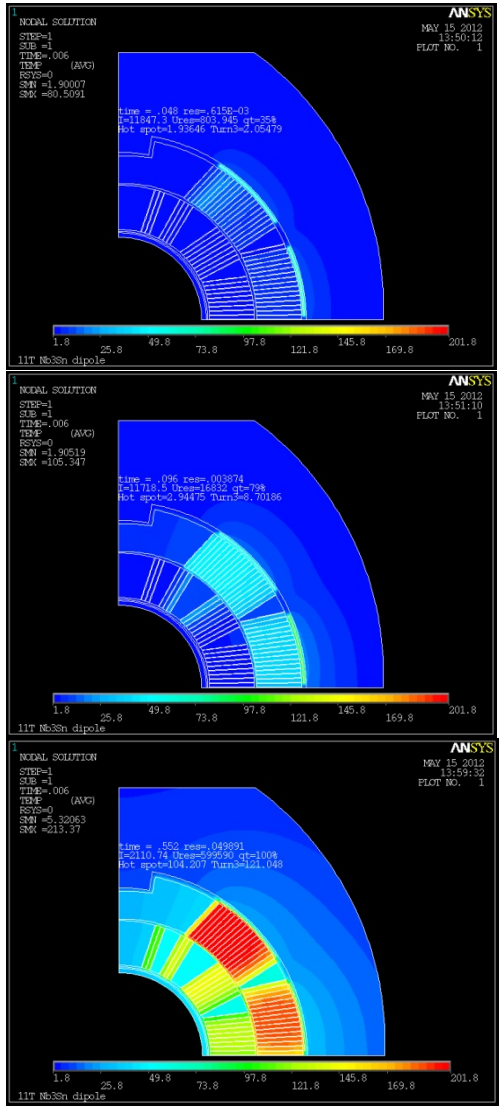

Figure 9: Temperature profile in the demonstrator magnet after 48 (top), 96 (middle) and 552 (bottom) ms from the heater induced quench.

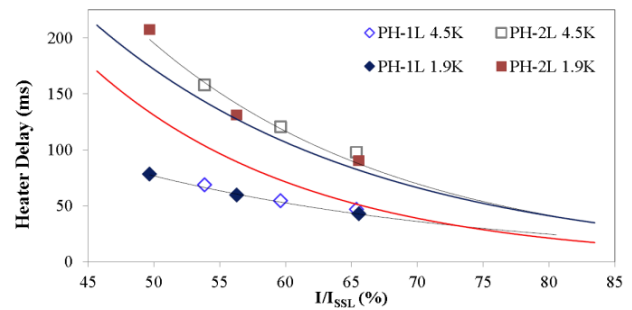

Figure 11: Estimated heater delay budget for operation with one (red line) or two (black line) heaters in each coil and measured heater delay at a different SSL ratio.

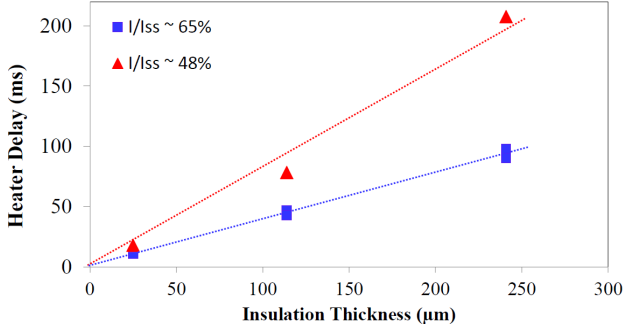

Figure 12: Heater delay time vs. insulation thickness.

*Effect of heater power and energy*

To study the additional possibilities to reduce the heater delay time and, thus, to increase the margin with respect to the total delay budget, the effects of the heater power and energy were measured. Heater delay time as a function of the peak heater power dissipated in the magnet at 4.5 K is shown in Fig. 13. The average peak heater power per heater area is defined as $I^2_{PH} R_{PH}/A$, where $I_{PH}$ is the maximum heater current (A), $R_{PH}$ and $A$ are the heater resistance (Ω) and area (cm$^2$) respectively. The data are shown at the magnet currents corresponding to 60% and 65% of its SSL at 4.5 K. Changing the heater power by almost a factor of two proportionally reduces the heater delay time for both heaters. The highest heater power density of 25 W/cm$^2$ was achieved during the test with the existed heater firing units.

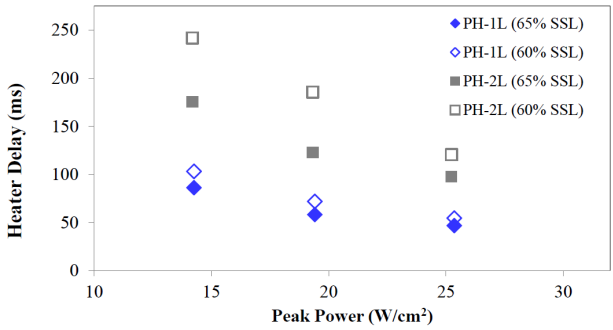

Figure 13: Heater delay as a function of peak dissipated power at 4.5 K.

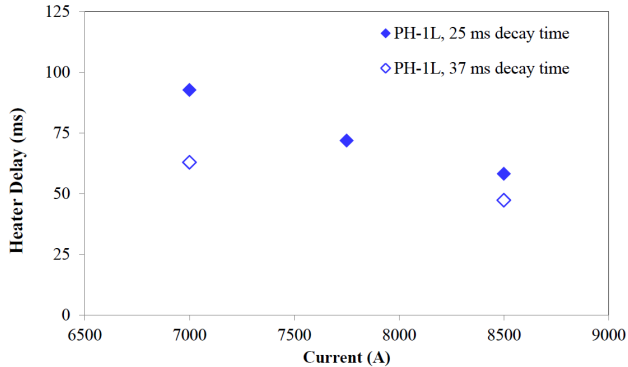

Figure 14: Heater delay as a function of magnet current for the peak heater power of ~ 20 W/cm$^2$ and different decay time constant of the heater circuit.

Heater delays could be further reduced by increasing the decay time constant (total energy deposited in heaters) of the heater circuit at the same peak heater power (Fig. 14).

*Quench development in low field and high field blocks*

Quench development and protection heater performance were studied for the Low Field (LF) and High Field (HF) outer-layer blocks since both these areas are covered by heaters. The heater strip width is not the same and as a consequence the peak power density is different in the LF and HF blocks.

The peak power density presented in the previous sub-section was averaged for both strips of the heater. The peak power density in the LF and HF areas can be presented as:

$$P_{LF} = 1.24 \cdot P_{av}, \qquad P_{HF} = P_{av}/1.24, \qquad (5)$$

where $P_{av}=I^2(R_{LF}+R_{HF})/(A_{LF}+A_{HF})$.

PH-1L and PH-2L heater delays in the LF and HF areas at 65% of SSL are shown in Fig. 15. The energy extraction circuit (dump) delay was 1 ms in these tests limiting possibilities of quench detection both in the HF and LF blocks. PH-1L heater delay in the low field area in most cases exceeded the quench detection time and thus the quench development in this area was not captured. That is why only once quench development was observed in the LF block for PH-1L with a delay time of ~20 ms with respect to the HF block.

Fig. 15 shows that all PH-2L induced quenches first developed in the low field area and only later in the high field area. The cause of this phenomenon is being investigated.

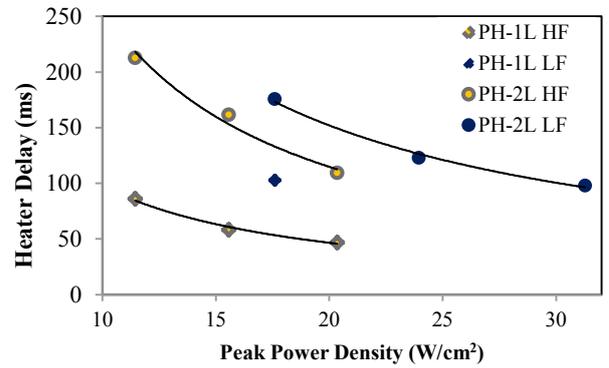

Figure 15: PH-1L and PH-2L heater delay in low and high field blocks as a function of peak dissipated power at 4.5 K.

However, this experiment shows that the delay between the HF and LF block quenches could be minimized or even completely avoided by optimizing the heater power in the HF and LF protection heaters.

Studies of LF and HF heater delay time will continue in next models. The protection heaters in the next 11 T dipole models will have only a single layer of Kapton insulation. The dump delay will be increased in order to investigate the quench development both in the low and high field blocks.

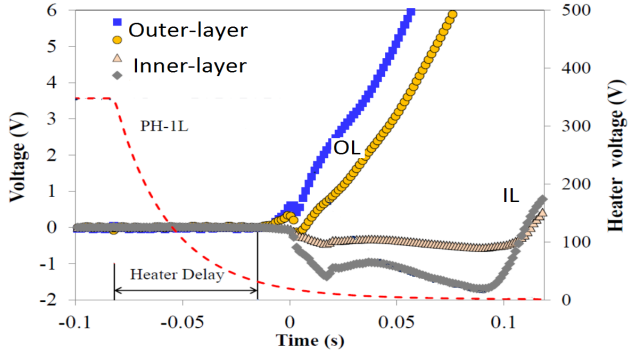

Figure 16: PH-1L heater-induced quench with a dump delay of 120 ms. Quench developed in 65 ms after heater ignition (PH-1L heater delay).

*Radial quench propagation*

To observe the quench propagation from the coil outer to the inner layer in heater-induced quenches at 4.5 K, the extraction dump was delayed by 120 ms. A quench at a magnet current of 8 kA (~62% of SSL) was provoked by igniting PH-1L while PH-2L was delayed and used for the magnet protection.

Figure 16 shows the development of the resistive voltage signal in the outer and inner coil layers. The heater voltage discharge in PH-1L is also shown in Fig. 16 (PH-2L ignition starts after the quench detection in the outer layer). After ~65 ms of the PH-1L ignition, a quench was initiated in the pole block of the outer coil layer. After an additional ~85 ms (still before the extraction dump was fired), clear resistive signals appeared in the inner coil layer segments. This experiment clearly confirms the rapid quench propagation from outer to inner layers in $Nb_3Sn$ accelerator magnets predicted by simulations in [16].

*Longitudinal Quench Propagation*

Most of the training quenches started in the mid-plane area of the outer coil layer and only a few quenches occurred in the inner-layer pole-turn segments with highest magnetic field [4]. The longitudinal quench propagation velocity was measured in one of the quenches in the inner-layer pole turn at 4.5 K using the time-of-flight method as ~27 m/s. Quench current in this ramp was 9440 A, which corresponds to 73% of SSL at 4.5 K.

The measured value of the longitudinal quench propagation velocity is comparable to, or higher than results obtained for other $Nb_3Sn$ magnets [19, 20]. Measurements of quench propagation velocity will continue on the next models with improved quench performance and coil instrumentation (spot heaters and additional voltage taps).

# EXTRAPOLATION TO LONG PROTOTYPE AND LHC CONDITIONS

To predict the efficiency of protection scheme with outer-layer heaters used in the 11 T dipole demonstrator under "LHC conditions", ROXIE quench protection module [21] and the LHC MB quench protection system parameters were used [22].

*ROXIE model calibration*

The ROXIE quench module uses a thermal network with one temperature node per half-turn in the cross-section. For heater simulations a 2D model was used. The heat propagates from turn to turn and from layer to layer through the insulation. Heaters are modeled as one temperature node per strip, with the associated heat capacity of a stainless steel strip. The electrical power is discharged into the heat capacity. The protection heater heats the coil turns under the heater, and, through the ground insulation, supplies heat to the helium bath at constant temperature.

In the model, the thermal conductivity between the heater and the coil, and between the heater and the helium bath, are determined from user-supplied thicknesses and insulation materials. The 0.125 mm glass-epoxy wrap around the coil is also taken into account. The model includes the quench-back effect with rather low inter-strand contact resistance in cable $R_c$=30 μΩ and $R_a$=0.3 μΩ. However, analysis shows that the corresponding quench-back effect reduces the coil maximum temperature only by 5% [22]. The model, however, does not include the thermal contact resistances between heater and Kapton, individual Kapton layers, and Kapton and coil or collars. To take into account these additional thermal resistances, scaling factors were used to tune the model using the experimental data. Another model shortcoming is that the heater is connected to an isothermal bath, rather than to the outer structure. As a consequence, in the case of low heater power and/or low currents, i.e., whenever heater delays are long, the heater cooling is too strong.

Model tuning was done to fit the heater delays measured at 1.9 K for PH-1L with a single layer of Kapton between heaters and coils. The results are shown in Fig. 17. The scaling factor for the thermal conductivity through the Kapton insulation used for tuning purposes for the single-layer case was set to 0.42. For completeness, the two-layer case was also modeled with a scaling factor of 0.33.

Using the updated ROXIE quench protection module the radial heat propagation time was also estimated. During the heater test [17] at 8000 A, with 350 V on a 9.6 mF capacitance of the heater power supply, the measured time delay between a first quench in the outer layer and a propagated quench in the inner layer was 85 ms (see Fig. 16). In a simulation with tuned ROXIE model, this delay was 110 ms which is also consistent with ANSYS model prediction calculated at 11.85 kA current (see Fig. 9). The results for some additional cases are presented in [22].

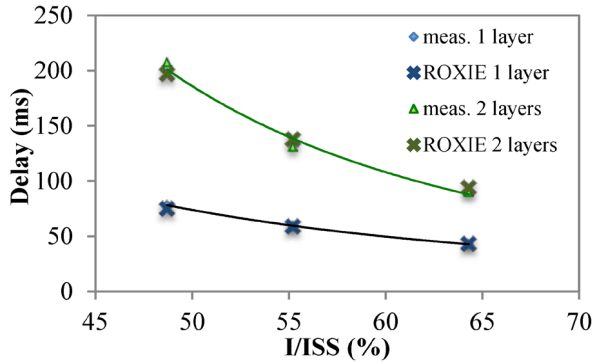

Figure 17: ROXIE model tuned to fit the measured heater delays.

*LHC Conditions*

Additional factors important for the 11 T dipole quench protection analysis in the LHC include the initial spread of the normal zone up to the detection threshold, validation time delays of the detection electronics, heater firing delays, the propagation of the normal zone into the inner layer, quench-back, and the number of turns under heaters to accelerate the current decay. Some of these parameters used for the LHC MBs are shown in Table 3.

Table 3: LHC MB quench-protection parameters

| Parameter | Value |
|---|---|
| Nominal detection threshold (V) | 0.1 |
| Nominal validation delay (ms) | 10 |
| Minimum heater-firing delay (ms) | 5 |
| Actual heater delay in RB circuits (ms) | <50 |

To estimate the time delay from the start of an initial 3 cm long resistive zone until the threshold voltage of 0.1 V is reached, the 3D thermal network model was implemented in ROXIE. Simulations yield 3 ms if the quench starts in the peak-field conductor (inner-layer) at nominal current, and 34 ms if the quench starts on the outer-layer midplane conductor at the nominal current. The simulated turn-to-turn delays in the respective locations were 3 ms and 22 ms. Longitudinal propagation velocities of ~29 m/s if the quench starts in the peak-field conductor and ~6 m/s if the quench starts on the outer-layer mid-plane which is consistent with the measured value of ~27 m/s in the demonstrator dipole (see Subsection "*Longitudinal quench propagation*"). These low values indicate that a finer discretization in the third dimension might be needed.

Using the calibrated ROXIE quench protection module and the above quench protection parameters, the efficiency of the outer-layer heaters used in the 2-m-long demonstrator dipole was estimated for realistic LHC conditions. Simulations were carried out using a 5.5-m-long single-aperture dipole magnet for two cases:

- Two protection heaters (LF and HF strips on both sides of each coil), 70 W/cm$^2$ maximum heater power and a time constant of 74 ms.
- Only one protection heater (one HF strip and one LF strip on the opposite side of the coil), the same maximum heater power and time constant.

The results of simulation for the two cases are summarized in Table 4. The analysis shows that the outer-layer protection heaters can keep the coil maximum temperature below 400 K with two operational heaters per coil. In the case with only one heater the calculated coil maximum temperature is reaching ~450 K.

Table 4: Quench simulations for the 11 T dipole under LHC conditions

| Parameter | 2 heaters | 1 heater |
|---|---|---|
| HF heater delay (ms) | 15 | 15 |
| LF heater delay (ms) | 28 | 28 |
| IL delay (ms) | 52 | 69 |
| QI total (MA$^2$s) | 16.5 | 18.6 |
| QI during current decay (MA$^2$s) | 11.5 | 13.6 |
| QI due to heater delay (MA$^2$s) | 2.1 | 2.1 |
| Peak coil temperature (K) | 378 | 456 |
| Peak heater temperature (K) | 292 | 292 |

Note that the above numerical model is a mix of optimistic and pessimistic assumptions. On the pessimistic side, the low quench propagation velocity increases the quench detection time and coil cooling in the model is underestimated (heat transfer to the helium bath, to the coil components such as wedges and poles, and to the mechanical structure). The ANSYS analysis shows that these effects play an important role in reducing the coil maximum temperature. On the optimistic side, the detection threshold is only 0.1 V with 10 ms validation delay, which will only work if the voltage spikes are short and few; also the heater-firing delay is set to that of the fastest systems in the current main dipole circuits. The model improvement and analysis of 11 T dipole protection under LHC conditions will continue.

## CONCLUSION

The high stored energy and low Cu/SC ratio in the cable, combined with the substantially larger temperature margins make the protection of the 11 T Nb$_3$Sn dipole a non-trivial problem.

Quench protection scheme based on the outer-layer protection heaters and two protection heater designs with 0.114 mm and 0.241 mm Kapton insulation thickness were analysed and experimentally evaluated for the 11 T Nb$_3$Sn dipole. The results of the study show acceptable heater efficiency and delay times for the heater with a single 0.114 mm thick Kapton film. This heater design will be used in the next 11 T dipole models. Fast quench propagation between the outer and inner coil layers was experimentally observed for the heater-induced quench. Longitudinal quench propagation velocity in a pole turn at ~73% of SSL was also measured. Due to limited magnet performance, heater tests were performed only at magnet currents up to 65% of SSL. Quench protection studies will continue with improved 11 T dipole models and coils with additional instrumentation.

The efficiency of the outer-layer protection heaters with 0.125 mm Kapton insulation to protect the 11 T dipole in LHC was also estimated using the improved ROXIE quench protection module, for both the regular case with two heaters and for only one heater per coil. The analysis shows that the outer-layer protection heaters can provide magnet protection (keep the coil maximum temperature below the limit of 400 K) in the nominal case with two operational heaters per coil. The calculated coil maximum temperature in the case with only one heater is 10% higher than the limit, reaching ~450 K. This case needs more study, both theoretical and experimental. However, the experimental data, obtained during the heater studies in 11 T dipole demonstrator, suggest that improvement of PH performance (reduction of the heater delay time) could be achieved by reducing the heater Kapton insulation thickness to 0.1 mm (~15%), and thermal contact resistances between Heater-Kapton-coil by gluing the heaters to the coil surface during coil impregnation. Some additional increase of the average peak heater power would also help.

Some general questions related to the quench protection of $Nb_3Sn$ accelerator magnets need to be further studied and addressed:

- What is the safe coil maximum temperature and average coil temperature under the heater for $Nb_3Sn$ magnets? Is $T_{max}$=400 K a safe limit? Is this limit universal or it depends on the magnet type and design?
- What is the role of longitudinal and transverse quench propagation, quench-back, coil cooling in protection of accelerator magnets?
- How will the radiation-hard insulation affect the magnet protection?
- Are the inner-layer and inter-layer protection heaters reliable? Can they be used for protection of $Nb_3Sn$ accelerator magnets? Are they compatible with the $Nb_3Sn$ magnet fabrication process and operation in superfluid helium?
- What is the effect of mechanical stress and mechanical shock during quench on the long-term magnet performance?

## ACKNOWLEDGMENT

Authors thank S.I. Bermudez and D. Tsirigkas (CERN) for their help with the ROXIE quench simulations.